\documentclass[aps,pra,groupedaddress,longbibliography,preprintnumbers,10pt,notitlepage]{revtex4-1}

\usepackage{graphicx,color}
\usepackage{epsfig}
\usepackage{dcolumn}
\usepackage{amsmath,amssymb,bm}
\usepackage{hyperref,url}
\usepackage[noabbrev]{cleveref}
\usepackage{CJK}
\usepackage{lineno}
\usepackage[english]{babel}

\usepackage{mathptmx} 

\providecommand\bnabla{\boldsymbol{\nabla}}
\providecommand\bcdot{\boldsymbol{\cdot}}
\newcommand{\tens}[1]{\bm{#1}}

\newcommand{\com}[1]{\textcolor{black}{#1}}

\newcommand{\tR}{t_\mathrm{R}}
\newcommand{\tRd}{t^*_\mathrm{R}}
\newcommand{\hmin}{h_\textrm{min}}
\newcommand{\hmind}{h^*_\textrm{min}}
\newcommand{\angstrom}{\mbox{\normalfont\AA}}

\begin{document}


\preprint{Accepted for publication in Phys. Rev. Fluids}

\title{Stokes theory of thin-film rupture}



\author{D. Moreno-Boza}
\email{damoreno@pa.uc3m.es}
\affiliation{Grupo de Mec\'anica de Fluidos, Departamento de Ingenier\'ia T\'ermica y de Fluidos, Universidad Carlos III de Madrid. Avda. de la Universidad 30, 28911, Legan\'es, Madrid, Spain.}

\author{A. Mart\'inez-Calvo}
\email{amcalvo@ing.uc3m.es}
\affiliation{Grupo de Mec\'anica de Fluidos, Departamento de Ingenier\'ia T\'ermica y de Fluidos, Universidad Carlos III de Madrid. Avda. de la Universidad 30, 28911, Legan\'es, Madrid, Spain.}

\author{A. Sevilla}
\email{alejandro.sevilla@uc3m.es}
\affiliation{Grupo de Mec\'anica de Fluidos, Departamento de Ingenier\'ia T\'ermica y de Fluidos, Universidad Carlos III de Madrid. Avda. de la Universidad 30, 28911, Legan\'es, Madrid, Spain.}


\begin{abstract}
The structure of the flow induced by the van der Waals destabilization of a non-wetting liquid film placed on a solid substrate is studied by means of theory and numerical simulations of the Stokes equations. Our analysis reveals that lubrication theory, which yields $\hmin\propto \tau^{1/5}$ where $\hmin$ is the minimum film thickness and $\tau$ is the time until breakup, cannot be used to describe the local flow close to rupture. Instead, the slender lubrication solution is shown to experience a crossover to a universal self-similar solution of the Stokes equations that yields $\hmin\propto \tau^{1/3}$, with an opening angle of $37^{\circ}$ off the solid.
\end{abstract}

\date{\today}

\maketitle

\section{Introduction}\label{sec:intro}

\com{Thin liquid films are ubiquitous in nature and everyday life, and they play important roles in many engineering processes, medical and physiological contexts, and in geophysics and biophysics, among other fields. As a consequence, many research efforts have been devoted to study the structure, stability and nonlinear dynamics of liquid films. For instance, falling films are relevant in coating processes~\cite{Huppert1982Nature,Kalliadasis2011}, gravity currents are important in geological phenomena ~\cite{Pattle1959,Huppert1982,Huppert1986}, and pulsed-laser-heated thin films are used in patterning and plasmonic applications~\cite{Makarov2016,Hughes2017,Kondic2019}. Tear-film dynamics~\cite{Braun2012,Fuller2012,Hermans2015,Dey2019} and surfactant replacement therapy~\cite{Jensen1992,Halpern1992,Jensen1993,Halpern1993,Grotberg1994,Halpern1998,Cassidy1999} are examples where the stability of surfactant-driven films is crucial for the healthy functioning of the eye and lungs. The reader is referred to~\cite{de1985wetting,oron1997long,bonn2009wetting,craster2009dynamics,Blossey2012,Kondic2019} for detailed and excellent reviews of the vast amount of scientific literature devoted to liquid films.}

\com{The present work is of fundamental character, and deals with} a non-wetting ultrathin liquid film placed on a solid substrate. \com{The liquid film} becomes unstable to infinitesimal surface waves when its thickness becomes smaller than about 100 nm, leading to a spinodal dewetting pathway coexistent with hole nucleation~\com{\cite{wyart1990drying,reiter1992dewetting,oron1997long,reiter1999destabilising,seemann2001dewetting,becker2003complex,thiele2007structure,craster2009dynamics,bonn2009wetting,Blossey2012}}. The spontaneous growth of perturbations takes place when the destabilizing van der Waals (vdW) forces exceed the stabilizing surface tension force, provided that the disturbance wavenumber is below a certain cut-off~\citep{scheludko1962certaines,Vrij1966,Ruckenstein1974}. Previous theoretical efforts to describe the nonlinear dynamics leading to film rupture were based on lubrication theory~\com{\citep{Williams1982,burelbach1988,ZhangLister1999,Blossey2012}}, which assumes that the longitudinal length scale is much larger than the film thickness, and provides models with simpler mathematical structure than the Navier-Stokes equations. Indeed, while the latter must be solved as a free boundary problem where the film thickness $h$ is part of the solution, the former leads to a partial differential equation (PDE) for $h$ as a function of the relevant spatial coordinates and time, that is physically transparent and more amenable to analysis.

In this paper, through dimensional arguments, numerical computations and similarity theory, we reveal that the Stokes flow close to the rupture singularity \com{is not slender, and} provides results markedly different from previous ones based on lubrication theory~\cite{ZhangLister1999}. Our local theory, motivated by dimensional analysis, is inspired by pioneering studies of singularities in free-surface flows without resorting to one-dimensional approximations of the equations of motion, in contexts like the breakup of liquid jets~\citep{Papageorgiou1995JFM,day1998self}, or the ejection of jets from Faraday waves~\citep{Lathrop2000}.

\section{Formulation}\label{sec:formulation}

\begin{figure}
    \centering
    \includegraphics[width=\textwidth]{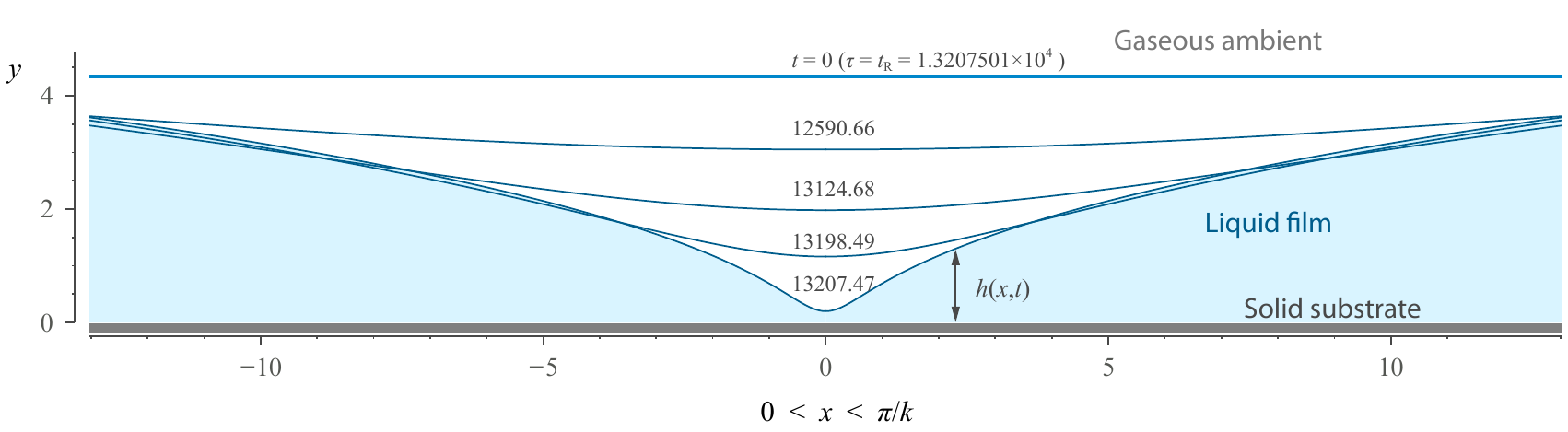}
    \caption{Schematics of the flow configuration, including a sample numerical evolution of the free surface for $h_o/a = 4.34$, where $a=[A/(6\pi\sigma)]^{1/2}$ is the molecular length scale~\cite{de1985wetting}. The free-surface shapes are plotted at the four time instants indicated with blue dots in Fig.~\ref{fig:fig2}$(b)$. The initial condition is $h_o/a(1-10^{-3}\cos{0.063\,x})$, with an associated rupture time $\tR \approx 1.3207501\times 10^{4}$ (Fig.~\ref{fig:fig2}$d$). \com{The entire computational domain, which spans a streamwise length $\pi / 0.063 \simeq 49.87$, is not represented for convenience.}}\label{fig:fig1} 
\end{figure}

Consider a planar film of Newtonian liquid of viscosity $\mu$ coating a solid surface that spans the $(x,z)$ plane. As sketched in Fig.~\ref{fig:fig1} the film, of initial height $h_o$, is surrounded by a passive gaseous atmosphere at constant pressure, such that the gas-liquid interface has a surface tension $\sigma$, and is described by the function $y = h(x,t)$. The liquid film, initially at rest, becomes unstable due to the long-range vdW forces, whose collective effects are modeled through a disjoining pressure, $A/(6 \pi h^3)$, with associated Hamaker constant $A$~\cite{Hamaker1937}, being $A>0$ in the non-wetting case considered herein. The latter intermolecular force model, which considers only non-retarded vdW interactions, is the simplest one among a hierarchy of existing models to rationalize the experimental observations~\cite{thiele2001dewetting,seemann2001dewetting,israelachvili2011intermolecular}. Note that, as argued in Appendix~A, liquid inertia is negligible under realistic experimental conditions. The cartesian components of the liquid velocity field in the $(x,y)$ directions are $(u,v)$, and the pressure field is $p$.

The molecular length scale~\cite{de1985wetting}, $a=[A/(6\pi\sigma)]^{1/2}$, is taken as the relevant characteristic length scale for the ultrathin liquid films considered in the present work, together with $\mu a/\sigma$, $\sigma/\mu$ and $A/(6\pi a^3)$ as time, velocity and pressure scales, respectively. The non-dimensional Stokes equations read
\begin{equation}
\label{eq:ns}
\bnabla \bcdot \bm{u} = 0, \quad
\bm{0} = - \bnabla \phi + \bnabla \bcdot \tens{T},
\end{equation}
where $\bm{u} = (u,v)$, $\tens{T} = - p\tens{I} + \bnabla \bm{u} + (\bnabla \bm{u})^{\rm{T}} $ is the liquid stress tensor and $\phi = h^{-3}$ is the dimensionless vdW potential. The accompanying boundary conditions include the non-slip condition $\bm{u} = 0$ at the solid wall $y = 0$, and 
\begin{eqnarray}
 \tens{T} \bcdot \bm{n} + \left(\bnabla \bcdot \bm{n}\right)\bm{n}&=& \bm{0}, \label{eq:stressBC}\\
 \bm{n} \bcdot \left({\partial}\bm{x}_s/{\partial}t-\bm{u}\right)&=&0, \label{eq:kinemBC}
\end{eqnarray} 
at the free surface $y = h(x,t)$, with corresponding parametrization $\bm{x}_s$ and unit normal vector $\bm{n}$, accounting for the stress balance and the kinematics of the interface, respectively.

In contrast with Stokes flow, the lubrication approximation provides the much simpler leading-order description 
\begin{equation}
h_t + \left(h^3 h_{xxx}/3 + h^{-1}h_x \right)_x=0,\label{eq:lubrication}
\end{equation}
governing the evolution of the free surface under the small-slope assumption~\cite{Williams1982,burelbach1988,ZhangLister1999}. Hereafter, subscripts will denote partial derivatives. Finally, note that the dimensionless initial film thickness, $h_o/a$, is the only governing parameter.

\section{Results}\label{sec:results}

\com{The presentation of results is divided into three parts. First, we will show that dimensional analysis demonstrates the existence of a non-slender self-similar solution of the first kind for the Stokes equations, governed by a balance between viscous and vdW forces, for which capillary forces are negligible. Second, we will present several comparisons of the film evolution given by the numerical integration of the Stokes equations, with that predicted by the lubrication model. Finally, we formulate and solve the self-similar Stokes problem, comparing the resulting solution with the rescaled numerical results for the full temporal evolution for times close to rupture.}

\subsection{Dimensional analysis}\label{subsec:diman}

Dimensional arguments suggest the existence of a similarity solution of the Stokes equations near film rupture that differs from the lubrication result~\cite{ZhangLister1999}. Indeed, the parametric dependences of the longitudinal velocity, transverse velocity, pressure and film thickness are
\begin{eqnarray}
\left[u^*, v^*, p^*\right]&=&\left[F_u^*,F_v^*,F_p^*\right](x^*,y^*,\tau_*,\mu,A,\sigma,h_o), \label{eq:funcuvpdim} \\
h^*&=&F_h^*(x^*,\tau_*,\mu,A,\sigma,h_o), \label{eq:funchdim}
\end{eqnarray}
where $\tau_*=\tRd-t^*$ is the time remaining to rupture, and asterisks denote the dimensional versions of the flow variables. Taking $(\tau_*,\mu,A)$ as dimensional basis, the Buckingham $\Pi$ theorem provides the reduced functional dependences,
\begin{eqnarray}
\left[u^*,v^*\right]&=&\left[\Pi_u,\Pi_v\right]\left(\xi,\eta,\Pi_{\sigma},\Pi_{h_o}\right)(A/\mu)^{1/3}\tau_*^{-2/3}, \label{eq:funcuvadim} \\
p^*&=&\Pi_p\left(\xi,\eta,\Pi_{\sigma},\Pi_{h_o}\right)\mu\tau_*^{-1}, \label{eq:funcpadim} \\
h^*&=&\Pi_h\left(\xi,\Pi_{\sigma},\Pi_{h_o}\right)(A/\mu)^{1/3}\tau_*^{1/3}, \label{eq:funchadim}
\end{eqnarray}
where $[\xi,\eta,\Pi_{h_o}]=(\mu/A)^{1/3}\tau_*^{-1/3}[x^*,y^*,h_o]$ and $\Pi_{\sigma}=\sigma/(A^{1/3}\mu^{2/3})\tau_*^{2/3}$.
When $\tau_*\to 0$, $\Pi_{\sigma}\to 0$ and $\Pi_{h_o}\to\infty$, suggesting that as rupture is approached surface tension forces become negligible, and that the local flow becomes independent of $h_o$. \com{Note also that the velocity components blow up near rupture, as dictated by~\eqref{eq:funcuvadim}, a common feature exhibited in most finite-time singularities in free surface flows.} Thus, we expect a local self-similar Stokes flow of the form
\begin{eqnarray}
 \left[u^*,v^*\right]&\xrightarrow{\tau_*\to 0}&\left[U\left(\xi,\eta\right),V\left(\xi,\eta\right)\right](A/\mu)^{1/3}\tau_*^{-2/3},\label{eq:diman_uv}\\
 p^*&\xrightarrow{\tau_*\to 0}&P\left(\xi,\eta\right)\mu/\tau_*,\label{eq:diman_p}\\
 h^*&\xrightarrow{\tau_*\to 0}&H\left(\xi\right)(A/\mu)^{1/3}\tau_*^{1/3}.\label{eq:diman_h}
\end{eqnarray}
\com{Note also that, since the self-similar scalings for $x$ and $y$ are the same, it can be anticipated that the \com{approach of the flow to the singularity} cannot be described using lubrication theory, which assumes that the characteristic length in the $x$ direction is much larger than the film thickness.}

\begin{figure*}[t]
    \centering
    \includegraphics[width=\textwidth]{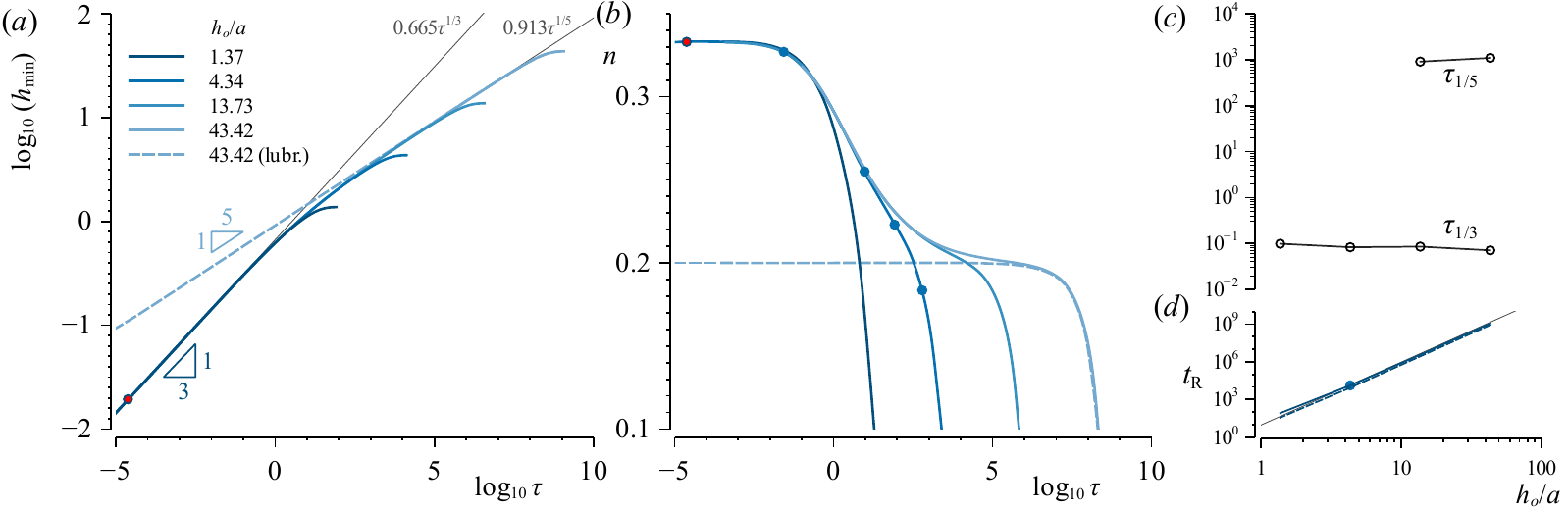}
    \caption{$(a)$ Minimum film thickness as a function of the time remaining to rupture (solid lines) for $h_o/a = (1.37, 4.34, 13.73, 43.42)$ and initial conditions $h_o/a(1-10^{-3}\cos{k\,x})$, using the corresponding optimal wavenumbers, $k = (0.17, 0.063, 0.021, 0.0065)$. The results obtained with the lubrication approximation~\cite{ZhangLister1999} are also presented for $h_o/a = 43.42$ (dashed line). $(b)$ Instantaneous exponent $n(\tau) = \mathrm{d} \log_{10}{\hmin} / \mathrm{d} \log_{10}{\tau}$. The blue dots indicate the times at which the interface profiles are plotted in Fig.~\ref{fig:fig1}, and the red dot marks the time chosen for the self-similarity test of Fig.~\ref{fig:fig4}. $(c)$ Times at which $n(\tau_{1/5})=1/5+0.1\times(1/3-1/5)$, and at which $n(\tau_{1/3})=1/5+0.9\times(1/3-1/5)$, obtained from the Stokes equations. $(d)$ Rupture time $\tR$ obtained from the Stokes equations (solid line), from the lubrication equation (dashed line), \com{and estimated from~\eqref{eq:tRLub} (thin line)}. The dot marks the rupture time associated with Fig.~\ref{fig:fig1}.}
    \label{fig:fig2}
\end{figure*}


\subsection{Flow evolution}\label{subsec:flowevol}

The equations~\eqref{eq:ns}--\eqref{eq:kinemBC} were numerically integrated with the \com{symmetry conditions $u=v_x=0$ imposed at the planes $x=0$ and $x=\pi/k$, and the initial conditions $\bm{u} = 0$ and $h(x,0)=h_o/a(1-10^{-3}\cos{kx})$ for $0 < x < \pi/k $. Here, the wavenumber} $k$ was chosen as the wavenumber of maximum amplification deduced from a linear stability analysis of Eqs.~\eqref{eq:ns}--\eqref{eq:kinemBC} \com{for the Stokes description~\cite{Jain1976, Gonzalez2016}, and of Eq.~\eqref{eq:lubrication} for the lubrication approximation~\cite{Vrij1966}, which yield, respectively, the dispersion relations
\begin{eqnarray}
\omega_{\text{ Stokes}} &=& \frac{3 \left(h_o/a\right)^{-4}- k^2}{ 2 k }\frac{ \sinh \left[2 (h_o/a) k\right]  - 2 (h_o/a) k }{ 1 + 2 (h_o/a)^2 k^2 +\cosh \left[2 (h_o/a) k\right]}, \label{eq:DRStokes} \\
\omega_{\text{ Lub}} &=& k^2\left[\frac{1}{(h_o/a)}-\frac{(h_o/a)^3}{3}k^2\right]. \label{eq:DRLub}
\end{eqnarray}
The Stokes dispersion relation~\eqref{eq:DRStokes} has the small-$k$ expansion 
\begin{equation}
\omega_{\text{ Stokes}}=k^2\left[\frac{1}{(h_o/a)}-\frac{(h_o/a)^3}{3}k^2\right]-\frac{9(h_o/a)}{5}k^4+\left[\frac{23(h_o/a)^3}{7}+\frac{3(h_o/a)^5}{5}\right]k^6+O\left(k^8\right),\label{eq:DRStokesSmallk}
\end{equation}
which, at $O(k^4)$, differs from the lubrication dispersion relation by the factor $9(h_o/a)k^4/5$. Note that the latter correction of the lubrication result is important for $h_o/a\sim 1$, but becomes negligible for $h_o/a \gg 1$. Using the lubrication equation~\eqref{eq:DRLub}, and the expansion~\eqref{eq:DRStokesSmallk} of the Stokes dispersion relation truncated at $O(k^6)$, the film rupture time can be estimated, using the corresponding maximum growth rates, as
\begin{eqnarray}
\tR^{\text{Lub}}&=& \frac{4}{3}\left(\frac{h_o}{a}\right)^5 \ln{(\varepsilon^{-1})},\label{eq:tRLub} \\
\frac{\tR^{\text{Stokes}}}{\tR^{\text{Lub}}}&=&1+\frac{27}{10}\left(\frac{h_o}{a}\right)^{-2}+O\left[\left(\frac{h_o}{a}\right)^{-4}\right] ,\label{eq:tRStokes}
\end{eqnarray}
respectively, where $\varepsilon$ is the initial disturbance amplitude. The approximation~\eqref{eq:tRLub} for $\tR$ is plotted as a thin line in Fig.~\ref{fig:fig2}($d$) for $\varepsilon=10^{-3}$, showing good agreement with the numerical rupture time. From Eq.~\eqref{eq:tRStokes} it is also deduced that the Stokes equations predict a slightly longer lifetime for the liquid film than the lubrication approximation}. A detailed description of the numerical techniques can be found in \com{Appendix~C}.

A representative numerical integration is presented in Fig.~\ref{fig:fig1} for $h_o/a=4.34$. The slightly disturbed flat film profile departs from the initial condition by virtue of the destabilizing vdW forces in a self-accelerated process, leading to a rupture singularity in a finite time $\tR$, whose precise computation involved an algebraic fitting procedure that took advantage of the anticipated power-law behavior $\hmin \propto \tau^{1/3}$ for $\tau \to 0$. Sample computations are depicted in Fig.~\ref{fig:fig2} for several values of $h_o/a$. The accompanying instantaneous exponent $n = \mathrm{d} \log_{10}{\hmin} / \mathrm{d} \log_{10}{\tau}$ reveals the persistent self-similar behavior $\hmin \to K_S\,\tau^{1/3}$ for $\tau \lesssim 0.1$ and all values of $h_o/a$, in agreement with dimensional analysis, where $K_S=0.665$. The solution of the lubrication equation~\eqref{eq:lubrication} for $h_o/a=43.42$ (dashed line) also exhibits a self-similar behavior~\cite{ZhangLister1999}, but with a different asymptotic law $\hmin \to K_L\,\tau^{1/5}$, where $K_L=0.913$. The crossover time between the lubrication and Stokes self-similar solutions can be estimated by equating $K_L\tau_c^{1/5}=K_S\tau_c^{1/3}\Rightarrow \tau_c=(K_L/K_S)^{15/2}=10.77$, with an associated minimum thickness $h_{\text{min},c}=K_S(K_L/K_S)^{5/2}=1.47$. Indeed, the results of Fig.~\ref{fig:fig2} for $h_o/a=43.42$ reveal that the evolution of $\hmin(\tau)$ obtained with the lubrication equation closely follows the Stokes result for $\tau \geq \tau_{1/5} \approx 10^3$ with a scaling exponent of $1/5$, followed by a long crossover for $\tau_{1/3} \leq \tau \leq \tau_{1/5}$, and finally reaching the $1/3$ power law for $\tau \leq \tau_{1/3}\approx 0.08$. In terms of the minimum film thickness, the $1/5$-scaling takes place for $\hmin\geq 3.63$, the corrossover for $0.29\leq \hmin \leq 3.63$, and the $1/3$-scaling for $\hmin\leq 0.29$. 

The failure of lubrication theory to predict the last stages of the rupture behavior observed in Fig.~\ref{fig:fig2} demands unraveling the local self-similar Stokes flow\com{, presented in the next section.}


\subsection{Self-similar solution}\label{subsec:similarity}

Dimensional analysis suggests substituting the similarity ansatz
\begin{equation}
    x = \tau^{1/3}\xi, \; y = \tau^{1/3} \eta, \; h = \tau^{1/3} f(\xi), \; u = \tau^{-2/3} U(\xi,\eta), \; v = \tau^{-2/3} V(\xi,\eta), \; p = \tau^{-1} P(\xi,\eta), \label{eq:ansatz}
\end{equation}
into~\eqref{eq:ns}--\eqref{eq:kinemBC} to elucidate the structure of the leading-order flow for $\tau \to 0$. The self-similar Stokes equations read
\begin{eqnarray}
 &&   U_\xi + V_\eta = 0, \label{eq:selfcont} \\
 &&   U_{\xi\xi} + U_{\eta\eta} = P_\xi - 3f^{-4}f_\xi , \label{eq:selfmomxi} \\
 &&   V_{\xi\xi} + V_{\eta\eta} = P_\eta, \label{eq:selfmometa}
\end{eqnarray} 
which must be integrated in $ 0 < \xi < \infty$, $0 < \eta < f(\xi)$, with the boundary conditions $U=V=0$ at the wall $\eta=0$, $U=V_\xi=0$ at the symmetry plane $\xi=0$, $0<\eta<f(0)$, and 
\begin{eqnarray}
 &&   (1+f_\xi^2) \, P - 2 V_\eta + 2 \, (V_\xi - f_\xi U_\xi + U_\eta) f_\xi = 0, \label{eq:selfnormstress} \\
 &&   (1-f_\xi^2) (V_\xi + U_\eta) + 2 \, (V_\eta - U_\xi) f_\xi = 0,  \label{eq:selftangstress} \\
 &&   f/3 - (\xi/3 + U) f_\xi + V = 0, \label{eq:selfkinem}
\end{eqnarray}  
at the unknown free surface $\eta = f(\xi)$. Notice that the leading-order contribution of the normal component of the stress balance~\eqref{eq:stressBC} is $O(\tau^{-1})$, while the capillary pressure is $O(\tau^{-1/3})$. Thus, as anticipated by dimensional arguments, surface tension does not contribute to the normal-stress equilibrium at leading order as $\tau\to 0$. The system of nonlinear elliptic PDEs~\eqref{eq:selfcont}--\eqref{eq:selfkinem} describing the local Stokes flow close to rupture is parameter-free. It is interesting to note that problems with similar mathematical structure appear in other free-surface flows like inertial focusing and jet breakup~\cite{Papageorgiou1995JFM,Lathrop2000}.

\begin{figure}
    \centering
    \includegraphics[width=0.6\textwidth]{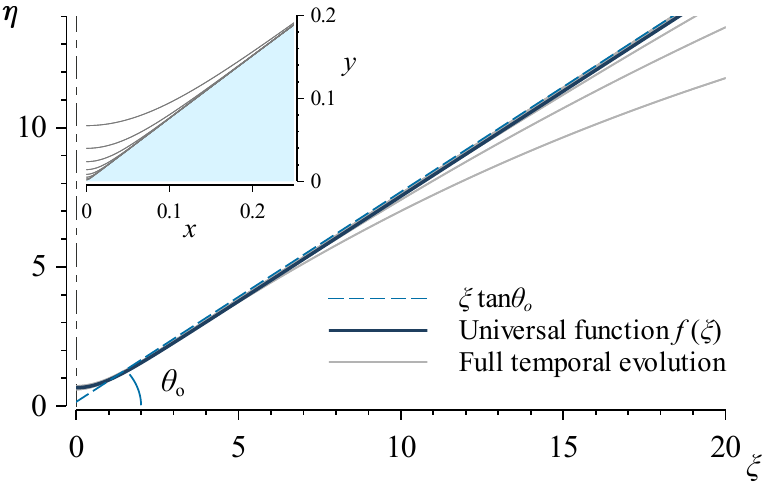}
    \caption{Comparison of the function $f(\xi)$ obtained from the self-similar solution (thick solid line) with eight rescaled profiles obtained from the numerical simulations for $h_o/a=4.34$ at times $-\log_{10} \tau = (2.98, 3.66, 4.33, 5.00,\ldots, 7.75$) (thin solid lines). The asymptotic shape $\xi \tan\theta_o$, with $\theta_o = 37^{\circ}$ (dashed line) is plotted with a vertical offset of $0.15$. The inset shows the unscaled profiles.}
    \label{fig:fig3}
\end{figure}

The asymptotic description of film rupture is completed by specifying the far-field boundary conditions at $\xi^2 + \eta^2 \gg 1$, $0 < \eta < f(\xi)$. Inspection of the kinematic boundary condition~\eqref{eq:selfkinem} reveals that $f\propto \xi$ for $\xi\gg 1$ if $U$ and $V$ are subdominant. This suggests that the shape of the free surface sufficiently far from the origin, $ r \gg 1$, is a wedge $\theta = \theta_o$, where $(r,\theta)$ are polar coordinates such that $\xi - \xi_o = r \cos \theta $ and $\eta = r \sin \theta$, and $(V_r,V_{\theta})$ are the associated radial and polar components of the velocity field. Insight of the far-field behavior was first obtained by numerically integrating~\eqref{eq:selfcont}--\eqref{eq:selfkinem} imposing a stress-free boundary condition sufficiently far from the origin, as explained in \com{Appendix~C, where a detailed description of the numerical technique employed to solve the similarity problem is provided}.

The examination of the radial and polar velocity profiles along rays $\theta = \textrm{constant}$ revealed that $V_r \sim V_{\theta} \sim r^{-(1+\lambda)}$ for $r \gg 1$, where $\lambda$ is a positive constant smaller than unity. This suggests a far-field stream function of the form~\cite{barenblatt1972self} $\psi = F(\theta)/r^{\lambda}$, such that $V_r = r^{-1} \partial\psi/\partial\theta = F' r^{-(1+\lambda)}$ and $V_{\theta} = - \partial \psi/\partial r = \lambda F r^{-(1+\lambda)}$. Since $\psi$ is biharmonic, $F$ is seen to be the solution to the fourth-order linear homogeneous equation 
\begin{equation}
    F^{\text{(iv)}} + \left[4 + 2\lambda (2+\lambda) \right] F'' + \lambda^2 (2 + \lambda)^2  F = 0,\label{eq:eqF}
\end{equation} 
with the no slip condition $F=F'=0$ at $\theta=0$, and the vanishing normal, $F'''+[4+3\lambda(2+\lambda)]F'=0$ and tangential, $F''- \lambda(2+\lambda)F=0$ stress boundary conditions at $\theta = \theta_o$. If $\lambda$ were known, Eq.~\eqref{eq:eqF} together with the boundary conditions discussed above would constitute a closed eigenvalue problem for the universal angle $\theta_o$. However, $\lambda$ is expected to be determined from the asymptotic matching with a near-field description for $\xi\ll 1$, which is beyond the scope of this work.

We propose the following alternative: nontrivial solutions to~\eqref{eq:eqF} satisfying the boundary conditions exist if
\begin{equation}
    \label{eq:lambdatheta0}
        (\lambda +1)^2 \cos (2 \theta_o )+\cos[2 \theta_o  (\lambda +1)] - \lambda  (\lambda +2) = 0,
\end{equation} 
which determines the pairs $(\theta_o,\lambda)$ classifying the family of allowed far-field solutions of the Stokes equations with a free-surface angle $\theta_o$. Thus one may i) extract $\theta_o$ from a numerical integration of~\eqref{eq:selfcont}--\eqref{eq:selfkinem} with a stress-free far-field boundary condition and then obtain $\lambda$ from \eqref{eq:lambdatheta0}, and ii) repeat the integration now imposing the far field variables entailed in the description of $F$. This iterative process converges very fast, and is stopped when the successive values of $\theta_o$ differ less than a prescribed tolerance\com{, providing the pair of values $(\theta_o,\lambda)=(37^{\circ},0.19)$.}

\begin{figure*}[t]
    \centering
    \includegraphics[width=\textwidth]{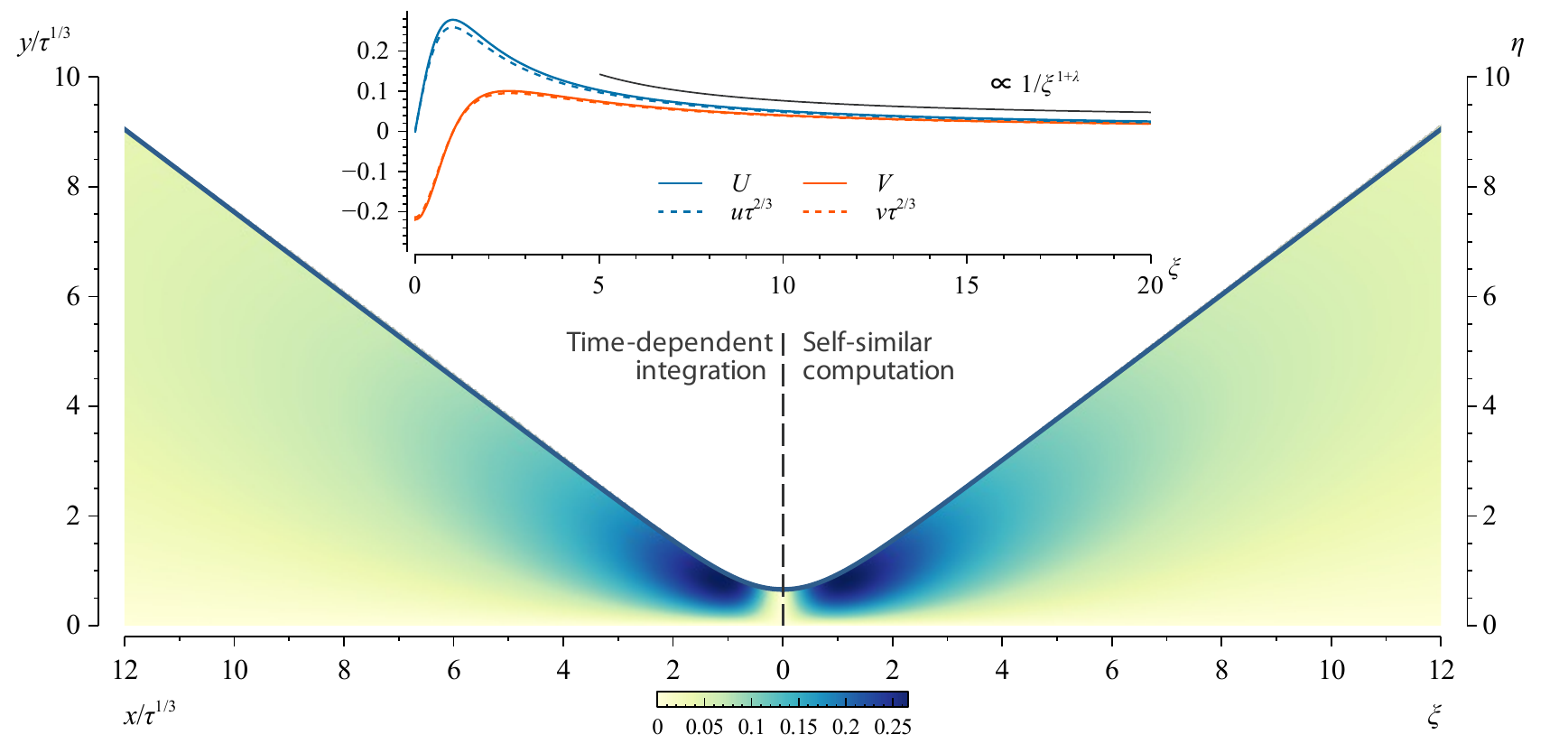}
    \caption{Isocontours of $U$ obtained by solving the self-similar system~\eqref{eq:selfcont}--\eqref{eq:selfkinem} (right panel), and by using the rescaled velocity $u\,\tau^{2/3}$ for $\log_{10}\tau = -4.62$ extracted from the simulation of~\eqref{eq:ns}--\eqref{eq:kinemBC} (left panel; red dot in Fig.~\ref{fig:fig1}). The inset shows the values of $U$ and $V$ along the interface, together with the asympotic far-field law $\xi^{-(1+\lambda)}$ for $\xi\gg 1$\com{, where $\lambda=0.19$ according to Eq.~\eqref{eq:lambdatheta0}}.}
    \label{fig:fig4}
\end{figure*}

The universal function $f(\xi)$ is represented in Fig.~\ref{fig:fig3} together with rescaled film profiles $h/\tau^{1/3}$ for $h_o/a = 4.34$, along with the far-field behavior $f = \xi \tan \theta_o$. The local expansion of the function $f$ for $\xi\ll 1$ has the form $f = f(0) + f''(0) \xi^2 + \mathcal{O}(\xi^4)$ due to the symmetry of the interface, with coefficients $f(0)=0.66$ and $f''(0) = 0.59$. As a final self-similarity test, Fig.~\ref{fig:fig4} displays isocontours of $U(\xi,\eta)$ obtained from the solution of~\eqref{eq:selfcont}-\eqref{eq:selfkinem} (right), and from the temporal integration of~\eqref{eq:ns}--\eqref{eq:kinemBC} for $\log_{10}{\tau}=-4.62$ (left), while the inset shows the corresponding profiles of $x$-- and $y$--velocity along the interface.

\section{Conclusions and future prospects}\label{sec:conclusions}

The self-similar solution obtained herein under Stokes flow arises from \com{a balance between viscous and van der Waals forces, while surface tension forces are asymptotically negligible as rupture is approached. Although this result is in agreement with the conclusions reached by~\citet{burelbach1988} for isothermal liquid films using lubrication theory, a more precise analysis by~\citet{ZhangLister1999} demonstrated that, in fact, the self-similar solution of the lubrication equation~\eqref{eq:lubrication} for $\tau\to 0$ arises from a balance between viscous, van der Waals and surface tension forces. In particular, it must be emphasized that the self-similar solution found by~\citet{ZhangLister1999} is inconsistent with the small-slope assumption, since the longitudinal length scales with time-to-rupture as $x\propto \tau^{2/5}$, while the minimum film thickness scales as $\hmin \propto \tau^{1/5}$. Thus, the slenderness of the film scales as $\hmin/x\propto \tau^{-1/5}\to \infty$ as $\tau\to 0$, invalidating the long-wavelength assumption near rupture. The latter inconsistency of the lubrication approach was already pointed out by~\citet{ZhangLister1999}, who estimated that the slenderness assumption breaks down when $\hmind\sim a$. Our analysis of the Stokes equations has revealed that, in fact, lubrication theory fails to describe the evolution of the film for $\hmind\lesssim 4a$. In particular, the shape of the free surface near rupture approaches a wedge $f \to \xi$ with a slope ${\rm d}f/d\xi=\tan{\theta_o}\approx 0.75$ for $\xi\gg 1$ according to our Stokes description, while $f \to \xi^{1/2}$ according to Eq.~\eqref{eq:lubrication}~\citep{ZhangLister1999}.} Although the leading-order lubrication theory does not describe the \com{last stages of the} vdW-induced rupture of thin films correctly, a higher-order theory might yield more accurate results~\cite{ruyer2000improved}. \com{Although  the lack of slenderness of the local Stokes solution may invalidate the use of higher-order lubrication theory, this line of research clearly deserves further inquiry}.

It is important to note that, regardless of the self-similar nature of the rupture dynamics, lubrication theory predicts an evolution for the liquid film that is markedly different from the Stokes description, as clearly evidenced by Fig.~\ref{fig:fig2}. Indeed, this difference appears during the early stages after the onset of the vdW instability, and increases over time. In particular, the 1/5 power law predicted by lubrication theory is only accomplished transiently during a very short intermediate time interval prior to the crossover to the 1/3 power law described here for the first time.

\com{Admittedly, the new self-similar solution presents a number of limitations, which we aim to identify below in terms of direct comparisons with experimental data available in the literature of thin films. Indeed,} the self-similar lubrication and Stokes regimes prevail for $\hmind > 3.63\,a$ and $\hmind < 0.29\,a$, respectively. Taking, for instance, $A = 2.2\times 10^{-20}$ J, $\sigma = 3.8 \times 10^{-2}$ Jm$^{-2}$ as representative values for ultrathin polymer films~\cite{becker2003complex} provides $a=1.75$ \angstrom. For $h_o = 4$ nm~\cite{becker2003complex}, the value of $h_o/a=22.82$ which, according to Fig.~\ref{fig:fig2}$(b)$, corresponds to a case where the self-similar lubrication regime is not established. The self-similar Stokes regime would be reached for $\hmind<0.5$ \angstrom, at which the continuum description is not valid. \com{ Additionally, taking the values $A = 3.90\times10^{-20}$ J, $\sigma = 2.07 \times 10^{-2}$ Jm$^{-2}$, given by~\cite{reiter1999destabilising} yield $a = 3.16$ \angstrom, which for a range of initial thicknesses $30$ nm $ < h_o < 110$ nm, provide values $95 < h_o/a < 348$. Although in the latter experiments the self-similar lubrication regime should be established, the computational cost quickly increases with increasing values of $h_o/a$. Although the numerical exploration of higher values of $h_o/a$ might prove useful to quantify the differences between the slender theory and the fully two-dimensional theory, it is expected that the evolution would follow the trends shown in Figs.~\ref{fig:fig2}(b)--(c) for the highest value of $h_o/a$. In particular, for $\tau > \tau_{1/5}$, both the lubrication and the Stokes evolutions would match before departing towards the 1/3-regime for intermediate crossover times $\tau_{1/3} < \tau < \tau_{1/5}$. More importantly, short-range intermolecular forces, not taken into account in the present analysis, would become important at larger values of $\hmind$. It is therefore concluded that, in the case of ultrathin polymer films, only the lubrication self-similar regime is experimentally realized, but only for values of $h_o/a\gtrsim 100$.}

\com{Liquefied metal films have been comparably less studied than polymer films, but there are a few experimental studies trying to characterize their instability and dynamics --see the review by~\cite{Kondic2019} and references therein. For instance, uniform thin films of Cu, Ag, and Ni of thicknesses 4 nm $< h_o <$ 10 nm, liquefied by laser irradiation over a solid substrate, are studied in~\cite{Gonzalez2013,Krishna2009,Mckeown2012}, respectively. Although there are still some uncertainties in the measurement of the Hamaker constant~\cite{Gonzalez2013,Gonzalez2016,Kondic2019}, the latter experimental studies provide the values $A^{Cu} = 2.58 \times 10^{-18}$ J, $A^{Ag} = 4.78 \times 10^{-18}$ J, and $A^{Ni} = 3.8 \times 10^{-18}$ J, which, considering $\sigma^{Cu} = 1.78$ Jm$^{-2}$, $\sigma^{Ag} = 0.93$ Jm$^{-2}$ and $\sigma^{Ni} = 1.71$ Jm$^{-2}$, yield $h_o/a^{Cu} = 14.43$, $h_o/a^{Ag} = 7.66$, and $h_o/a^{Ni} = 11.65$, for a film thickness of $h_o = 4$ nm. According to the findings presented herein, lubrication theory would provide an incorrect description for these three particular cases, as evidenced by figure~\ref{fig:fig2}($b$). The self-similar Stokes regime would be reached for $h_{\min}^{*,Co}<0.8$ \angstrom, $h_{\min}^{*,Ag}<1.51$ \angstrom, and $h_{\min}^{*,Ni}<1.0$ \angstrom, respectively, thereby compromising the continuum approximation. More importantly, as discussed in detail by~\cite{Gonzalez2016,Kondic2019}, in the case of liquefied metal films the liquid inertia can become relevant during film drainage and rupture. The latter fact will surely alter the self-similar structure obtained herein using the Stokes equations, as already shown by~\cite{Garg2017} for liquid films with power-law rheology.}

\com{An alternative and promising theoretical approach to characterize the flow of thin films at the nanoscale is performing molecular-dynamics (MD) simulations~\cite{Fowlkes2012,Nguyen2012,Nguyen2014}. In particular, it was shown in~\cite{Nguyen2012} that MD simulations and the weak-slip lubrication model~\cite{Munch2005,Fetzer2005} yield almost identical results in different geometries and flow configurations of metal liquid films when the thickness is about $10$ nm. In particular, these authors stressed that the slip length is essential to obtain consistent predictions between both frameworks. Moreover,~\cite{Moseler2000,Eggers2002} showed, in the case of liquid nanojets, that the addition of thermal fluctuations to the deterministic lubrication equation is crucial to obtain results in agreement with MD simulations. As for thin liquid films, there are a few theoretical and computational studies comparing the linear and nonlinear dynamics of the stochastic lubrication approximation with its deterministic counterpart~\cite{davidovitch2005spreading,grun2006thin,fetzer2007thermal,nesic2015fully}. However, the implication of thermally triggered fluctuations on the linear and nonlinear regimes within the Stokes framework remains as an open problem, to be compared with results obtained from the stochastic lubrication theory and MD simulations.}

\com{Another poorly understood aspect that deserves further study is the characteristic lifetime of the ultrathin film. According to Eq.~\eqref{eq:tRStokes}, the Stokes equations predict slightly longer rupture times than lubrication theory. However, as far as we know, a systematic and rigorous comparison of the theoretically predicted lifetimes with the experimental observations is still missing. Indeed, a mismatch between the rupture times between the experiments and the numerical simulations of the lubrication equation in polymer films was reported in~\cite{becker2003complex}. However, it remains unclear if this disagreement may be attributed to thermal noise~\cite{mecke2005thermal,grun2006thin,fetzer2007thermal}, to slippage effects~\cite{kargupta2004instability,Nguyen2012}, or to a poor characterization of the intermolecular interactions~\cite{pahlavan2018thin}.}

Natural extensions of our work include for instance the effect of wall slip~\citep{peschka2019signatures}, the study of axisymmetric rupture~\citep{ZhangLister1999,witelski2000dynamics}, the breakup of free films~\cite{vrij1968rupture,vaynblat2001rupture,champougny2017break}, the influence of surfactants~\cite{Edwards1995,warner2002dewetting}, liquid-liquid dewetting~\cite{wyart1993liquid,pototsky2004alternative,ward2011interfacial}, the influence of polymer rheology~\cite{Blossey2012}, and the effect of thermal noise~\cite{davidovitch2005spreading,grun2006thin,fetzer2007thermal,nesic2015fully}. \com{Inertial effects should also be considered, as required to account for the dynamics of liquid metal films~\citep{Gonzalez2013,Gonzalez2016, Kondic2019}}. \com{Finally, of} particular importance is the inclusion of more detailed models of intermolecular interactions~\cite{seemann2001dewetting,thiele2001dewetting,israelachvili2011intermolecular,pahlavan2018thin}, as required to account for the resulting dewetting patterns and their long-term coarsening~\cite{becker2003complex,glasner2003coarsening}.


\begin{acknowledgments}
The authors thank J. Rivero-Rodr\'iguez and B. Scheid for key numerical advice, A.L. S\'anchez, C. Mart\'inez-Baz\'an and J.M. Gordillo for their enduring teaching and encouragement, W. Coenen for carefully reading the manuscript, and H.A. Stone for sharing his insights, and for his kind hospitality at the Complex Fluids Group, Princeton. This research was funded by the Spanish MINECO, Subdirecci\'on General de Gesti\'on de Ayudas a la Investigaci\'on, through project DPI2015-71901-REDT, and by the Spanish MCIU-Agencia Estatal de Investigaci\'on through project DPI2017-88201-C3-3-R, partly financed through FEDER European funds. A.M.-C. also acknowledges support from the Spanish MECD through the grant FPU16/02562 and to its associated program Ayudas a la Movilidad 2018 during his stay at Princeton University.
\end{acknowledgments}

\appendix

\section{Inertial effects}
\label{sec:intertialeffects}
\com{In this appendix we briefly assess the accuracy of the assumption of negligible inertial effects}. The relative importance of liquid inertia compared with the viscous forces is given by the local Reynolds number
\begin{equation}
Re = \frac{\rho h^*_{\text{min}}}{\mu}\frac{{\rm d}h^*_{\text{min}}}{{\rm d}\tau^*}\Rightarrow 
Re=\frac{\rho a \sigma}{\mu^2}\,\hmin\frac{{\rm d}\hmin}{{\rm d}\tau},\label{eq:localRe1} 
\end{equation}
where $\rho$ is the liquid density and $\rho a \sigma/\mu^2$ is the Laplace number based on the liquid properties and the molecular length scale $a=[A/(6\pi\sigma)]^{1/2}$. Since $\hmin = 0.665\,\tau^{1/3}$ for $\tau\to 0$, ${\rm d}\hmin/{\rm d}\tau = 0.222\,\tau^{-2/3}$, and the local Reynolds number,
\begin{equation}
Re = 0.034\,\frac{\rho\sqrt{A\sigma}}{\mu^2}\,\tau^{-1/3}\to \infty\quad\text{as}\quad \tau\to 0.\label{eq:localRe2} 
\end{equation}
However, the inertial crossover time, $\tau_I$, defined by the condition $Re(\tau_I)=1$, accomplishes
\begin{equation}
 \tau_I = 3.9\times 10^{-5}\,\frac{\rho^3\left(A\sigma\right)^{3/2}}{\mu^6}. \label{eq:inertial_crossover}
\end{equation}
Taking $\rho=10^3$ kg$\,$m$^{-3}$, $A = 10^{-19}$ J, $\sigma=0.03$ Jm$^{-2}$, $\mu=1$ kg$\,$m$^{-1}\,$s$^{-1}$ provides an inertial crossover time of $\tau_I \approx 1.6\times 10^{-22}$, at a minimum radius $0.665\,\tau_I^{1/3}\approx 3.6\times 10^{-8}$, which is eight orders of magnitude smaller than the molecular length scale. Similar conclusions where obtained by~\citet{ZhangLister1999} in their lubrication analysis. 

It is thereby deduced that \com{inertia} can be safely neglected in the analysis of the van der Waals instability of thin \com{polymer} liquid films. \com{ By way of contrast, it should be noted that in situations concerning the thinning of liquid metal films, care must be taken in retaining the inertial effects, as discussed in~\cite{Gonzalez2013,Gonzalez2016, Kondic2019}. For instance, the global Reynolds numbers considered in the second reference scale up to $\sim 10^4$, to be compared with that from~\cite{becker2003complex}, on the order of $10^{-15}$, for which the estimations given in the previous paragraph are representative.}

\section{ Near-rupture velocities }

\com{Let us now provide an estimation of the typical velocities occurring near rupture. To that end, we take the fluid properties given in~\ref{sec:intertialeffects}, and a value of $\tau$ at which the short-range repulsive forces balance the long-range attractive vdW forces. The latter balance requires introducing a more realistic version of the potential, not considered in the main text, which according to~\citep{becker2003complex} may be written in dimensionless form as 
\begin{equation} 
\label{eq:completepotential}
\phi(h) = h^{-3} - \frac{ 48 \pi \epsilon }{A a^6} \, h^{-9}, 
\end{equation} 
where the second term includes the strength of the short-range part of the potential, namely, $\epsilon = 6.3 \times 10^{-76} \, \textrm{Jm}^6$. An estimation of the dimensionless precursor film thickness would be given by balancing both terms in~\eqref{eq:completepotential} resulting in $h_\textrm{prec} = \frac{1}{a} (48 \pi \epsilon/A)^{1/6}$, which yields a value $h_\textrm{prec} \simeq 2.36$. According to~\ref{fig:fig2}(a), this value of $h$ would lie well within the crossover region between the 1$/$5 and the 1$/$3 regimes, although the associated value of $\tau$ would hardly be modified. For instance, assuming the validity of the 1/3 power-law, the corresponding value of $\tau \simeq 44.6$ which yields dimensionless velocities on the order of $0.08$ according to~\eqref{eq:funcuvadim} assuming $O(1)$ values of the self-similar variables $U$ and $V$. Recasting these into dimensional form provides velocities on the order of $0.08 \sigma/\mu = 2.4 \times 10^{-3} \, \textrm{m}/\textrm{s}$.}

\section{Numerical techniques}

\com{In this appendix we describe the numerical techniques employed to solve the Stokes equations~\eqref{eq:ns}, the lubrication equation~\eqref{eq:lubrication}, and the self-similar Stokes problem~\eqref{eq:selfcont}--\eqref{eq:selfkinem}.} All the equations were written in weak form \com{in terms  of the corresponding inner products}, upon convenient use of Green's identities, rendering them amenable for the use of finite elements for the spatial discretization. {\sc Comsol}~\cite{COMSOL1998} was the software of choice for the implementation of such numerical techniques.

\subsection{The Stokes equations}

\begin{figure}
    \centering
    \includegraphics[width=0.56\textwidth]{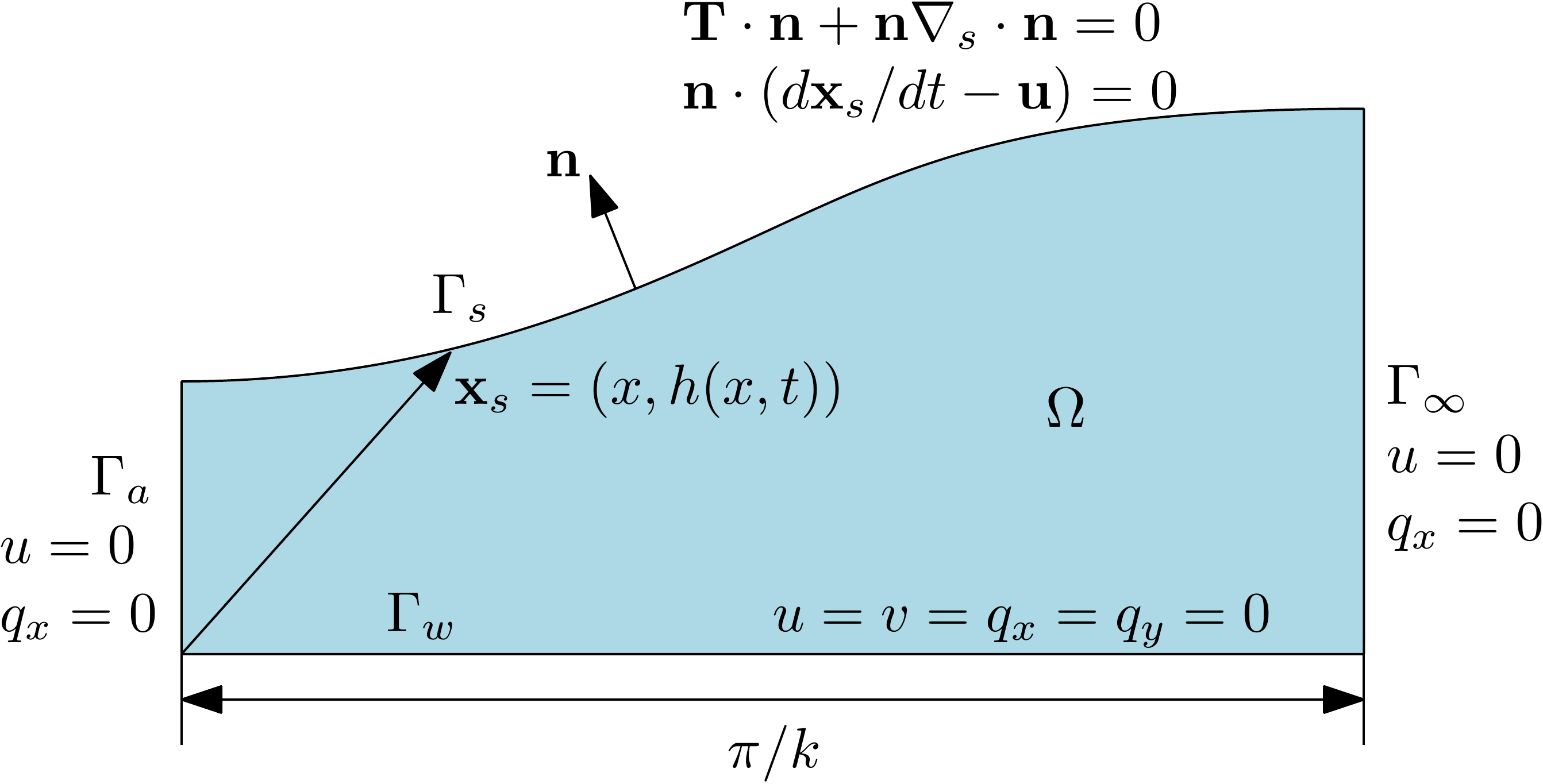}
    \hspace{0.4cm}
    \includegraphics[width=0.36\textwidth]{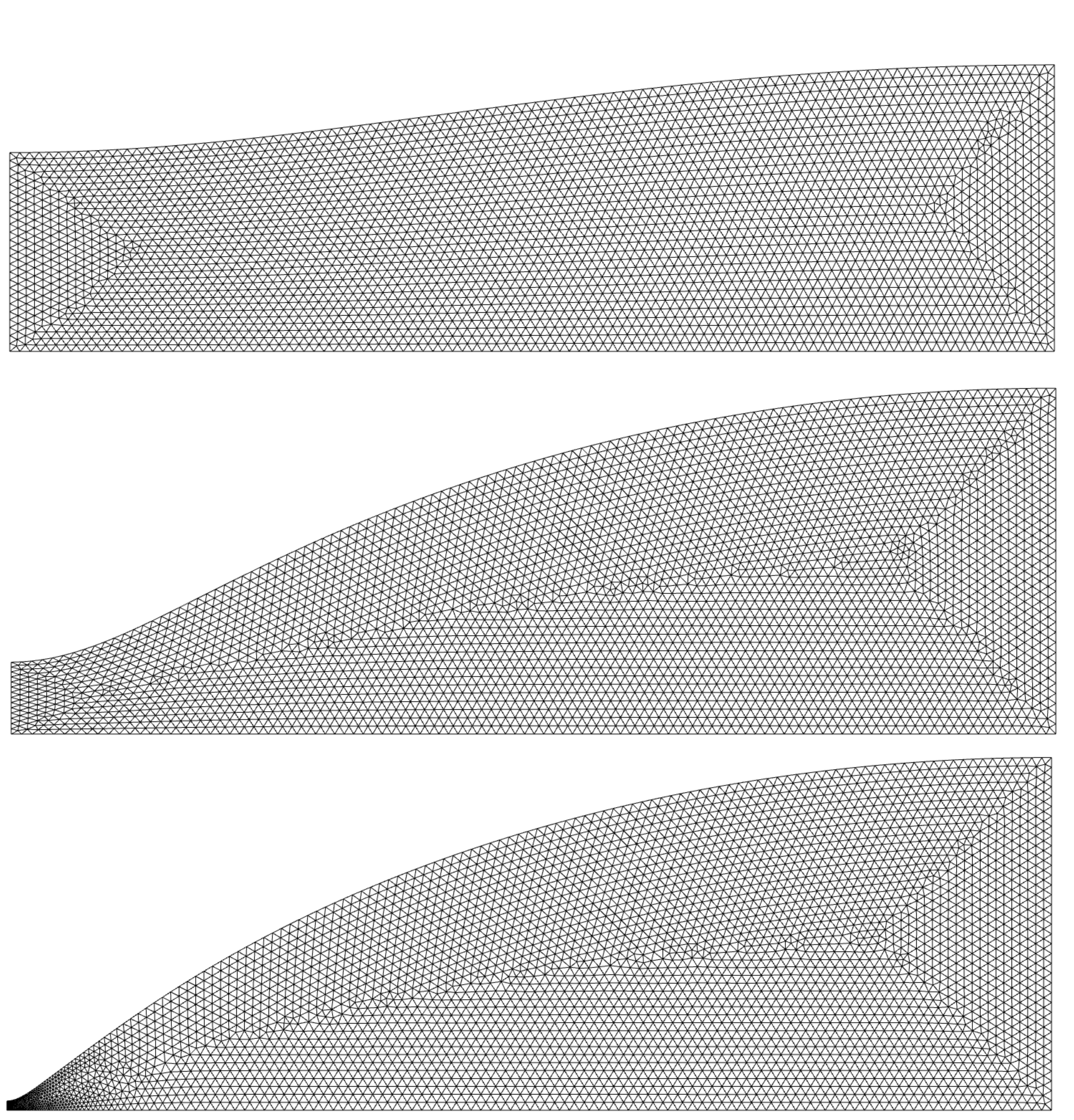}
    \caption{Left panel: sketch of the geometry including the boundary conditions for the velocity and mesh displacement fields. Right panel: sample deformed meshes for $h_o/a = 1.37$ and $t = (72.766, 82.376, 82.617)$ in the upper, middle and lower panels, respectively.}\label{fig:sketch_stokes}
\end{figure}

The dimensionless Stokes equations were written in weak form using suitable test functions for the velocity and pressure, namely $\tilde{\bm{u}}$ and $\tilde{p}$, as follows:
\begin{equation}
    \label{eq:nsweak}
    \int_\Omega \left[ \tilde{p} \nabla \cdot \bm{u} -  (p + \phi) \nabla \cdot \tilde{\bm{u}} + \left(\nabla \bm{u} + \nabla \bm{u}^\mathrm{T} \right) : \nabla \tilde{\bm{u}}  \right] \, \mathrm{d} \Omega + \int^{}_{\Gamma_s} \nabla_s \cdot \tilde{\bm{u}} \, \mathrm{d}S  = 0,
\end{equation} where $\Omega$ is the deformable computational domain bounded by $\Gamma_s \cup \Gamma_w \cup \Gamma_a \cup \Gamma_\infty$, the surface gradient operator is defined as $\nabla_s = (\tens{I} - \bm{n}\bm{n}) \cdot \nabla$, $\mathrm{d} \Omega$ is the volume element and $\mathrm{d} S$ is the surface element. The remaining boundary conditions are specified in the sketch of Fig.~\ref{fig:sketch_stokes}.

Equation~\eqref{eq:nsweak} was discretized using Taylor-Hood triangular elements for pressure and velocity (and the corresponding test functions) to ensure numerical stability. The use of an arbitrary Lagrangian-Eulerian (ALE) technique for the tracking of the interface allowed us to impose the kinematic boundary condition along $\Gamma_s$ by prescribing the normal velocity of the mesh. The displacement of the mesh elements was computed by solving a Laplace equation for the displacement field, namely $\nabla^2 \bm{q} = 0$, $\bm{q} = (q_x,q_y)$, with suitable boundary conditions.

The temporal discretization of the system of nonlinear equations was carried out using a variable-step BDF method with $2/4$ variable order. The relative tolerance of the nonlinear solver was set to $10^{-7}$. The initial conditions corresponded to a quiescent state $u = v = 0$ and a perturbed interface $h(x,0) = h_o/a\left(1 - 10^{-3} \cos{kx}\right)$. The time-dependent solver was complemented with an automatic remeshing algorithm which redistributed the mesh elements when the deformation of the domain become sufficiently large. The numerical limitations of the computational techniques employed in this study precluded the minimum film thickness $\hmin = h(0,t)$ to decrease below $\approx 10^{-4}$.

\subsection{The lubrication equation}

The lubrication equation for the temporal evolution of the film thickness~\com{\eqref{eq:lubrication}, which is herein written again for convenience},
\begin{equation}
h_t + \left(h^3 h_{xxx}/3 + h^{-1} h_x\right)_x=0, \quad t > 0, \quad 0 < x < \pi  / k,
\end{equation} 
\com{needs} to be integrated with boundary conditions $h_x(0,t) = h_x(\pi/k,t) = 0$, and initial condition $h(x,0) = \frac{10}{9}\frac{h_o}{a}\left(1 - \frac{1}{10}\cos{k x}\right)$, as used in~\cite{ZhangLister1999}, which was integrated by recasting it into the conservative second-order form 
\begin{equation}
    \label{eqG}
    (h_t,0) + \partial_x \bm{G} = \bm{g}, \quad  \text{with} \quad \bm{G} = \left[ h_x/h + h^3 h_{xxx}/3, \,  h_x \right], \quad \bm{g} = (0,h_{xxx}),
\end{equation} 
and discretized using quadratic Lagrange elements over a non-uniform partition of the domain with maximum element size of $10^{-7}$ close to the origin $x = 0$. The time-stepper algorithm was the same as that used for the Stokes equations, namely a $2/4$-BDF.

\begin{figure}
    \centering
    \includegraphics[width = 0.8\textwidth]{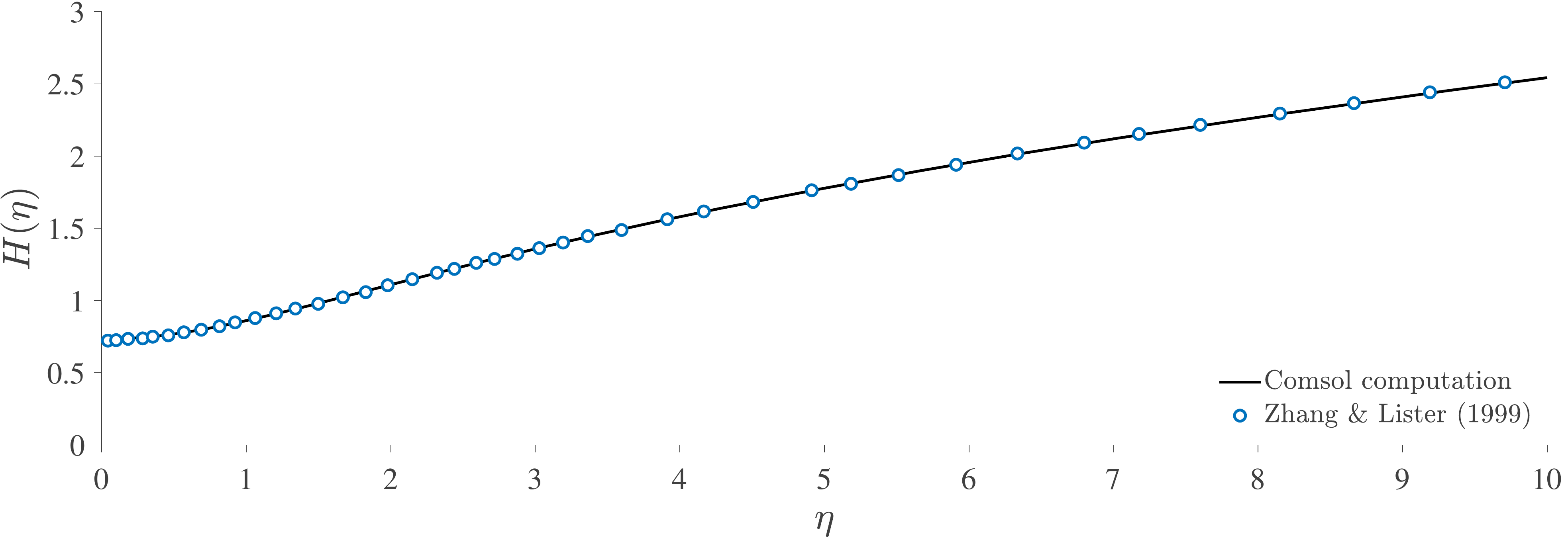}
    \caption{Comparison with the self-similar solution $H(\eta)$ computed by~\citet{ZhangLister1999} and the numerical integration of the lubrication equation performed with {\sc Comsol}~\cite{COMSOL1998}.}
    \label{fig:complister}
\end{figure}

The self-similar scaling of the film shape $h/(\tR - t)^{1/5} = H(\eta)$ with $\eta = x/(\tR - t)^{1/5}$ (note that this $\eta$ is different from that appearing in the main text) recovered the values obtained by~\citet{ZhangLister1999}, namely $H(0) = 0.7327$ and $H''(0) = 0.3010$ as seen in Fig.~\ref{fig:complister}, which served as a benchmark for our numerical scheme.

\begin{figure}
    \centering
    \includegraphics[width = \textwidth]{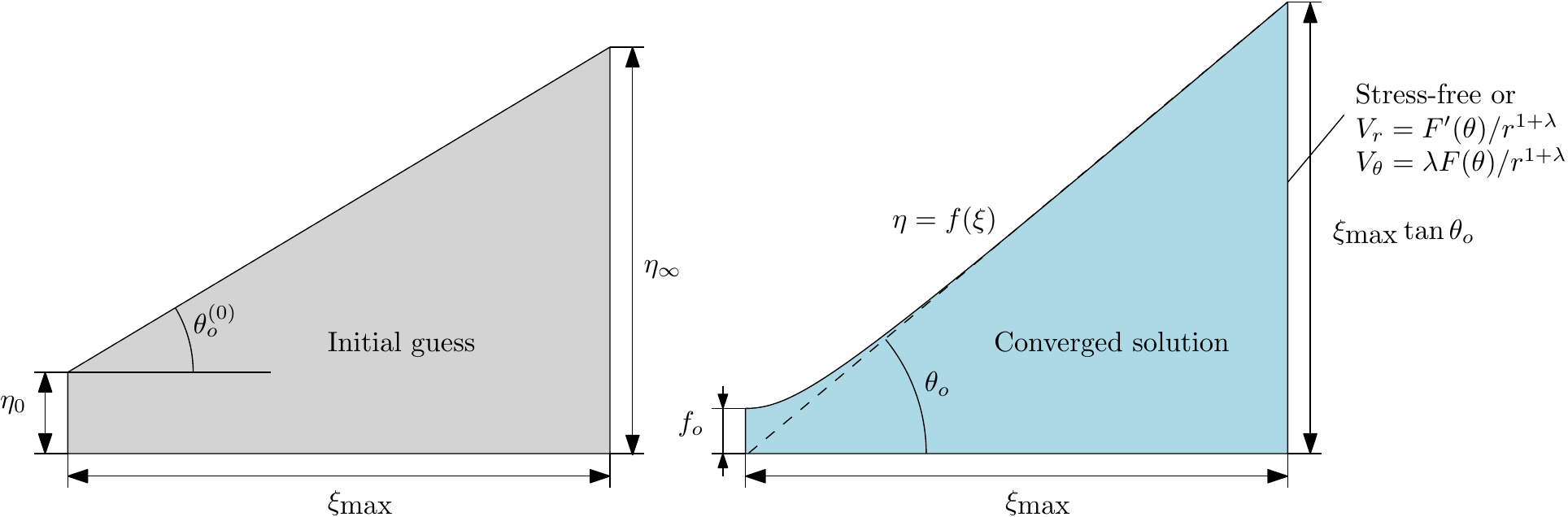}
    \caption{Left panel: sketch of the initial geometry used as initial guess for the Newton--Raphson algorithm. Right panel: final solution obtained after the root-finding iterations. In particular $f(\xi)$ and $\theta_o$ are obtained from the calculation. The boundary conditions at $\xi^2 + \eta^2 \gg 1$ and $0 < \eta < f(\xi)$ are also represented.}
    \label{fig:sketch_self}
\end{figure}

\subsection{The self-similar Stokes problem}

The elliptic system of PDEs for the self-similar variables $U$, $V$, $P$ and the a priori-unknown interface shape $f(\xi)$, was discretized using similar techniques as those employed for the complete Stokes equations. The associated weak form is
\begin{equation}
    \label{eq:selfsimweak}
    \int_\Omega \left[ \tilde{P} \nabla \cdot \bm{U} - \left[P + f^{-3} \right] \nabla \cdot \tilde{\bm{U}} + \left(\nabla \bm{U} + \nabla \bm{U}^\mathrm{T} \right) : \nabla \tilde{\bm{U}} \right] \, \mathrm{d} \Omega = 0 ,
\end{equation} 
where $\bm{U} = U \bm{e}_\xi + V \bm{e}_\eta$ and the del operator acts on the similarity coordinates $\xi$ and $\eta$. The boundary condition along the unknown free surface is that of vanishing stresses
\begin{equation}
    \left(-P \tens{I} + \nabla \bm{U} + \nabla \bm{U}^\mathrm{T} \right) \cdot \bm{n}   = 0,
\end{equation} 
which is naturally accommodated in weak form. The ALE method was used again in this time-independent computation but this time with an extra degree of freedom, namely, the vertical displacement $\delta$ of the free surface $\Gamma_s$. This displacement was leveraged as a Lagrange multiplier for the imposition of the kinematic boundary condition by solving the weak boundary PDE
\begin{equation}
    \int_{\Gamma_s} \tilde{\delta} \, \left[ f/3 - (\xi/3 + U) f_{\xi} + V \right] \, \mathrm{d}\Gamma = 0,
\end{equation} 
where $\tilde{\delta}$ is the test function for the vertical displacement $\delta$ discretized using first-order Lagrange elements and $\mathrm{d}\Gamma$ is the line element along $\Gamma_s$. This additional equation enabled us to use a standard Newton--Raphson root-finding algorithm to iteratively solve for all the unknowns upon a tolerance, fixed to $10^{-5}$, provided a suitable initial guess. Such an initial guess is depicted in Fig.~\ref{fig:sketch_self}, comprising a quadrilateral computational domain of vertical dimensions $\eta_0$ and $\eta_\infty$, horizontal span $\xi_{\max}$ and initial angle $\theta_o^{(0)}$ and vanishing velocities $U = V = 0$. A convenient choice for the tentative angle and $\eta_0$ was provided by the scaled film shapes obtained from the full temporal evolution of the Stokes equations near the breakup singularity. With this choice of the numerical parameters, the iterative algorithm described in the main text converged in a few iterations.




%

\end{document}